\documentclass[amstex,useAMS,epsf,amssymb,usenatbib,twocolumn]{mn2e}
\usepackage{graphicx}
\usepackage{epsfig}
\usepackage{amssymb,amsmath,color}
\usepackage{hyperref}
\usepackage{soul}
\sethlcolor{white}
\usepackage[normalem]{ulem}
\usepackage{url}
\usepackage{breakurl}
\usepackage{tabularx}
\usepackage{natbib}

\definecolor{lightblue}{rgb}{.90,.90,1}
\definecolor{lightorange}{rgb}{0.996, 0.847, 0.694}
\definecolor{lightgreen}{rgb}{0.82,0.94,0.75}
\definecolor{lightviolet}{rgb}{0.86, 0.82, 1}
\definecolor{turq}{rgb}{0.68, 0.89, 0.87}

\newcommand{\fermilat}{{\it Fermi}-LAT}
\newcommand{\gray}{$\gamma$-ray}
\newcommand{\grays}{$\gamma$ rays}

\def\apj{ApJ}                 
\def\apjl{ApJ}                
\def\apjs{ApJS}               
\def\aap{A\&A}                
\def\aapr{A\&A~Rev.}          
\def\mnras{MNRAS}             

\def\nat{Nature}              

\title[Fermi Galactic transients]{{An update on Fermi-LAT transients in the Galactic plane, including strong activity of Cygnus X-3 in mid-2020}}

\author[D. A. Prokhorov, A. Moraghan]{D. A. Prokhorov$^{1}$\thanks{E-mail:d.prokhorov@uva.nl},
A. Moraghan$^2$\thanks{E-mail:ajm@asiaa.sinica.edu.tw} \\
~\\
$^{1}$ GRAPPA, Anton Pannekoek Institute for Astronomy, University of Amsterdam, Science Park 904, 1098
XH Amsterdam, The Netherlands
\\
$^{2}$ Academia Sinica Institute of Astronomy and Astrophysics, 11F
of AS/NTU Astronomy-Mathematics Building, No.1, Sec. 4, Roosevelt Rd.,\\
Taipei 10617, Taiwan}

\date{\today}

\pagerange{\pageref{firstpage}--\pageref{lastpage}} \pubyear{2022}

\begin{document}

\maketitle

\begin{abstract}
We present a search for Galactic transient \gray{} sources using 13 years of the \textit{Fermi} Large Area Telescope data. The search is based on a recently developed variable-size sliding-time-window (VSSTW) analysis and aimed at studying variable $\gamma$-ray emission from binary systems, 
including novae, $\gamma$-ray binaries, and microquasars.
Compared to the previous search for transient
sources at random positions in the sky with 11.5 years of data, we included $\gamma$ rays with energies down to 500 MeV, increased a number of test positions, and extended the data set by adding data collected between February 2020 and July 2021. 
These refinements allowed us to detect additional three novae, V1324 Sco, V5855 Sgr, V357 Mus,
and one \gray{} binary, PSR B1259-63, with the VSSTW method. 
Our search revealed a \gray{} flare from the microquasar, Cygnus X-3, occurred in 2020. 
When applied to equal quarters of the data, the analysis provided us with detections of repeating
signals from PSR B1259-63, LS I +61$^{\circ}$303, PSR J2021+4026, and Cygnus X-3. 
While the Cygnus X-3 was bright in $\gamma$ rays in mid-2020, it was in a soft X-ray state and 
we found that its \gray{} emission was modulated with the orbital period.

\end{abstract}


\section{Introduction}

\

Transients in the Milky Way consist of a variety of sources.
Gamma-ray emitting binaries include four classes of $\gamma$-ray sources: 
($\mathrm{i}$) so-called $\gamma$-ray binaries; high-mass X-ray binary systems 
whose spectral energy distribution peaks at energies above MeV,
($\mathrm{ii}$) novae; thermonuclear explosions in binaries following accretion on to
white dwarfs, ($\mathrm{iii}$) microquasars; binaries powered by accretion on to a compact object
that display relativistic jets, and ($\mathrm{iv}$) colliding wind binaries; binaries
in which stellar outflows develop shocks giving rise to $\gamma$-ray emission. 
There are a variety of physical processes  responsible for the $\gamma$-ray emission for these 
different source classes \citep[for a review, see][]{Dubus, Paredes2019}. 
The detections of binaries in the GeV \gray{} band were based on the data 
collected by
the Large Area Telescope \citep[LAT;][]{Atwood2009} on-board the \textit{Fermi} 
Gamma-ray Space Telescope launched in June 2008 and we highlight some notable sources 
as follows:
\begin{itemize}
\item \textbf{Gamma-ray binaries.} During the first years of the \textit{Fermi} mission, 
three \gray{} binaries, LS I +61${^{\circ}}$303, LS 5039, and 1FGL J1018.6-5856, were detected 
\citep[][]{ls2009b, ls2009a, 1fglbin}.
In these papers, the \gray{} binaries were identified on the basis of their modulated  emission
with periods from a few to several days, see also the blind all-sky search for cyclic $\gamma$-ray sources by \citet[][]{PM2017}
and the most recently detected \gray{} binary, 4FGL J1405.1-6119 \citep[][]{Corbet2019}. 
In addition, \gray{} binaries, PSR B1259-63 and HESS J0632+057, 
with periods longer than a hundred days were detected by \citet[]{Tam2011, Caliandro2015} 
and by \citet[][]{Li2017}, respectively. Long-term GeV \gray{} variability of LS I +61${^{\circ}}$303 
with a superorbital period of 1667 days was reported by \citet[][]{superorb} and \citet[][]{PMV2021}. 
\item \textbf{Novae.} \textit{Fermi}-LAT detected 17 Galactic novae.
The first of these sources, V407 Cyg, was observed in March 2010 \citep{V407}. 
The brightest novae, such as V407 Cyg 2010 and V1324 Sco 2012, produce \gray{} fluxes about 10 times higher than the faintest ones, such as V549 Vel \citep[see,][]{LiKwanLok2020, Franckowiak2018}.
Multi-wavelength observations of novae are useful for their identification
and their discoveries are made in the optical band.
The duration of \gray{} emission from a nova is about two weeks \citep[e.g.,][]{Franckowiak2018}.
\item \textbf{Microquasars.} The first microquasar detected with \textit{Fermi}-LAT, 
Cygnus X-3, produced \gray{} flares in 2008 and 2009 \citep[][]{CygX3Sci}.
These \gray{} flaring events of Cygnus X-3 were also detected with AGILE \citep[][]{CygX3Nat}. The 
\gray{} flares of the source were observed during a soft X-ray state and allowed the \textit{Fermi}-LAT collaboration to establish that the corresponding \gray{} flux was modulated with a period of 
$4.7917\pm0.0011$ hours \citep[][]{CygX3Sci}, which is ascribed to the orbital period of the binary system. 
Another microquasar, Cygnus X-1, was detected at GeV energies
during hard X-ray states \citep[][]{Zanin2016}. 
Gamma-ray studies of 
other microquasars, including SS 433 and V404 Cyg, in the GeV band
are an active research field \citep[e.g.,][]{SS433, V404}.
\item \textbf{Colliding wind binaries.} Gamma-ray emitters belonging to this class include $\eta$ Carinae \citep[][]{etaCar2010} and $\gamma^2$ Velorum \citep[][]{Pshirkov2016}.
A \textit{Fermi}-LAT analysis of two full orbits and the third periastron of $\eta$ Carinae 
and hints of \gray{} orbital variability from $\gamma^2$ Velorum
were reported by \citet[][]{MartiDevesa2021} and \citet[][]{MartiDevesa2020}, respectively. 
\end{itemize}

Other sources in the Milky Way also produce transient $\gamma$-ray signals. Among these sources are transitional millisecond pulsars switching from accretion to a radio pulsar stage, e.g., a low-mass X-ray binary transition system PSR J1023+0038 \citep[][]{Stappers}, and young pulsar wind nebulae, e.g. the Crab nebula \citep[][]{CrabAGILE, CrabScience}. In addition, flaring, nearby, young M-dwarf stars are potential transient $\gamma$-ray sources in the Milky Way, see \citet[][]{Ohm2018} and the discussion by \citet[][]{Loh2017} on a high Galactic latitude transient event detected at a position consistent with DG CVn, but likely associated with a flaring background blazar. 

The \textit{Fermi}-LAT provides unprecedented sensitivity for all-sky monitoring of \gray{}
activity. Analysis techniques applied to searches for transient sources have different levels 
of detail and coverage. Searches for variable \gray{} emission at different positions
inside the large region of the
sky, e.g., the Galactic plane \citep[][]{Neronov12} or the entire
sky \citep[the \textit{Fermi} all-sky variability
analysis by][]{Fermi13, Fermi17}, on the time scale of weeks or months
use a measure of variability computed as, e.g., the maximum
deviation of the flux from the average value. The reduced $\chi^{2}$
of the fit of the light curve with the constant flux is another technique
which is adopted in the \textit{Fermi}-LAT catalogue
\citep[][]{Fermi20} for testing about 5,000 \gray{} sources. Both
these statistics allow tests of a large number of positions or
sources and are not computationally expensive.
However, these techniques have a predetermined time interval at which
variability is searched.
In \citet[][]{PMV2021}, the authors developed a variable-size sliding-time-window (VSSTW)
technique and applied it, in addition to a search for \gray{} emission from
supernovae, to a search for transient \gray{} sources
at random positions in the sky. The search by means of 
the VSSTW technique accounts for a start and 
duration of emission which serve as two variables. The VSSTW method allowed the authors to confirm 
the presence of transient \gray{} emission from transitional pulsars, solar flares, \gray{} bursts, novae, and the Crab Nebula,
and was successful in finding both short (e.g. solar flares) and long (e.g. transitional pulsars) 
high-\gray{}-flux states of sources.

The most recent search for transient \gray{} sources with full coverage of the Galactic plane 
was based on 7.4 years of \textit{Fermi}-LAT data \citep[][]{Fermi17}. 
Given that \textit{Fermi}-LAT accumulated 13 years of data by August 2021, a new search 
was warranted. 
In \citet[][]{PMV2021},
the authors performed an all-sky VSSTW test search for transients in \textit{Fermi}-LAT data
and found a new strong transient \gray{} source projected onto the Galactic plane, that is near PSR
J0205+6449 and flared in 2017. In addition to this new source, \citet[][]{PMV2021} confirmed  
7 flares of novae and the superorbital modulation of \gray{} emission from LS I +61${^{\circ}}$303. 
Since the area of the sky probabilistically covered by that initial search 
is $\approx$63\% (that is $(1-1/\mathrm{e})\times100\%$) of the sky, there are known transient 
sources that escaped detection in that test search, notably a bright nova V1324 Sco 2012.
To increase the number of transient \gray{} sources and in particular
\gray{} emitting binaries detected with the VSSTW technique in the Galactic plane, we performed a new search using more \textit{Fermi}-LAT data as well as full coverage and 
finer pixelization of the Galactic plane.

\section{Observations and analysis}

In this section, we describe \textit{Fermi}-LAT observations, data reduction, and the VSSTW analysis.

\subsection{\textit{Fermi}-LAT observations and data reduction}

\fermilat{} is a pair-conversion telescope and has been scanning the sky continuously since August 2008.
Due to its large detector area and field of view ($\approx 20\%$ of the sky), \fermilat{} allows efficient monitoring of the $\gamma$-ray sky.
The telescope provides an angular resolution per single event of $1.5^{\circ}$ at 0.5 GeV, 
$1.0^{\circ}$ at 0.8 GeV, narrowing to $0.5^{\circ}$ at 2 GeV, and further narrowing to $0.1^{\circ}$ above 10 GeV.
At energies below $\sim$10 GeV, the accuracy of the directional reconstruction of $\gamma$ rays is limited by multiple scattering in the converter foils of \fermilat{}. 
Given both the angular resolution dependence with energy and the broadband sensitivity to sources with power-law spectra\footnote{\burl{http://www.slac.stanford.edu/exp/glast/groups/canda/lat\_Performance.htm}}, 
we selected the optimal lower energy limit of 0.5 GeV. We made this selection with two purposes: to tighten the point spread function (PSF) and
to include $\gamma$ rays with energies between 0.5 GeV and 0.8 GeV. The latter allows a more detailed study of $\gamma$-ray sources with soft spectra, such as Cygnus X-3 \citep[][]{CygX3Sci} and PSR B1259-63 \cite[][]{Tam2011, psrb12592020}, than in \citet[][]{PMV2021}. We selected the upper energy limit of 500 GeV, because of the small amount of detected \gray{} events above this energy.

We downloaded the \fermilat{} Pass 8 (P8R3) data from the Fermi Science Support Center consisting of
680 weeks of the \texttt{SOURCE} class data (evtype=128), collected between 2008-08-04 and 2021-08-12 and including 80
weeks of data additional to the data set analysed by \citet[][]{PMV2021}. 
We performed the data analysis using the \texttt{FERMITOOLS} v1.2.23 package. We rejected \gray{} events with
zenith angles larger than $90^{\circ}$ to reduce contamination by
albedo \grays{} from the Earth and applied the recommended cuts on
the data quality (DATA\_QUAL$>0$ \&\& LAT\_CONFIG$==1$). We binned
the data into time intervals of one week and in four energy bands,
that are 0.5-0.8 GeV, 0.8-2.0 GeV, 2.0-5.0 GeV, and 5.0-500.0 GeV. 
Treatment of these four energy bands provided us with an analysis independent on the photon index.
We further binned the \fermilat{} events using the HEALPIX package\footnote{\burl{{http://healpix.sourceforge.net}}} \citep[][]{Gorski} into a map
of resolution $N_\mathrm{side}$ = 512 in Galactic coordinates with
`RING' pixel ordering. With these settings, the total number of
pixels is equal to $12\times512^2$=3145728 and the area of each pixel is
$4\pi\times(180/\pi)^2/(12\times512^2)$=0.0131 deg$^2$, given that HEALPix subdivides the sphere into 12 equal-area base pixels at the lowest resolution. To compute the exposure, we used the standard tools
\texttt{gtltcube} and \texttt{gtexpcube2}. To correct the livetime
for the zenith angle cut, we used the `zmax' option on the command line.
For this analysis, we selected 68608 positions whose centers coincide with those of the HEALPix grid of resolution $N_\mathrm{side}$ = 256 in Galactic coordinates with `RING' pixel ordering and are between $-5^{\circ}$ and $+5^{\circ}$ Galactic latitudes. On top of that, we excluded $1\fdg5$ radius regions 
around the Geminga and Vela pulsars since systematic errors, which are not taken into 
account in our analysis, might exceed statistical ones for these two 
brightest $\gamma$-ray sources \citep[][]{Geminga}. 
We used events within a 0\fdg5 radius circle centered on each of the remaining 68476 positions.
Between $-5^{\circ}$ and $+5^{\circ}$ Galactic latitudes the number of $\gamma$-ray sources in the 4FGL-DR2 catalogue \citep[][]{Fermi20} is 1101 and larger than the number of regions with the radius of the PSF 68\% containment in the lowest energy band, which is $(360^{\circ}\times10^{\circ}/(\pi\times1\fdg5\times1\fdg5))=509$.
The 0\fdg5 radius aperture covers an area twice larger than the aperture used in \citet[][]{PMV2021} and
sufficient to accumulate a significant number of events from the potential source
in the four energy bands, but still small enough to suppress the contamination from blazars with variable $\gamma$-ray fluxes or other neighbouring sources, especially in the two lowest energy bands. 

\subsection{Time sliding window search}

Using the publicly available python code that employs a likelihood analysis \citep[for a review of likelihood theory, see][]{Mattox}
for finding the most statistically significant time interval of a
high flux at a given position in the sky \footnote{\burl{https://zenodo.org/record/4739389}} developed by \citet[][]{PMV2021}, we performed a search for transient \gray{} sources in the 
Galactic plane. Following their approach, we compared a model with the presence of a temporary bright state
above a steady flux level and a model assuming a source with a steady
flux in time. For this purpose, we computed a likelihood for each of these models taking into 
account the numbers of expected and observed counts from the source during each week and in each energy band, 
and multiplying Poisson probabilities for all weeks.
To evaluate the significance of evidence for a bright state, we used the Test Statistic (TS), 
which is defined as twice the difference in log-likelihood maximums of two models.
Since the two considered models are nested, the probability distribution of the TS is
approximately a chi-square distribution with four degrees of
freedom (one degree for each energy band) accordingly to Wilks' theorem \citep[][]{Wilks}.
In this search, a sliding-time window can start at any of the 680 weeks and have any
duration unless the 680th week is reached. 

The number of weeks is 680 and the number of uncorrelated 
positions tested in the Galactic plane is high. We estimated that the average 
value of a statistical level for signals in a cleaned sample is 3.5$\sigma$ corresponding to
a trial factor of 2457 due to a choice of time interval. We estimated  
the number of uncorrelated test positions as $3600/(\pi\times2^{2})=287$, where 3600 deg$^{2}$ is the
total area covered by the region of the Galactic plane selected for this analysis and 
the radius of 2$^{\circ}$ minimises a position correlation of signals. 
We adapted a global significance level
where we indicated the significance level after taking the ``look elsewhere effect''
into account. This effect
is quantified in terms of a trial factor that is the ratio of
the probability of observing the excess in the obtained time interval
to the probability of observing it with the same local significance level
anywhere in the allowed range at any of the uncorrelated test positions.
Accounting for the trial factor, that is $2457\times287\approx7.0e5$, 
a global significance level above 5.0$\sigma$ translates to a local significance level 
higher than 7.0$\sigma$. The criterion for classifying a high flux time interval at a given position 
as a transient signal satisfies the convention of a $5\sigma$ global significance level 
being required to qualify as a discovery. 

The performed time sliding window search showed that apart from the two brightest sources, 
the Vela and Geminga pulsars, systematic uncertainties should be taken into account for another bright $\gamma$-ray source, 
PSR J1826-1256, previously discussed by \citet[][]{Neronov12}, \citet[][]{Fermi17}, and \citet[][]{PMV2021}. 
We checked and found that 
the classification of this source as a transient source is not robust if systematic uncertainties in exposures are present at a level of $3\%$. 

We also found that some high flux time intervals with a global significance above 5.0$\sigma$ 
correspond to \gray{} flares of known blazars from the Fermi-LAT catalogue \citep[][]{Fermi20}. 
Both blazars located close to and at a distance larger than $0\fdg5$ from a tested position can affect the search,
e.g., NRAO 676 which is $\simeq0\fdg1$ from a tested position and PKS 1830-21, 
located at a Galactic latitude of -5.7$^{\circ}$, which
is $\simeq0\fdg8$ from another tested position correspond to highly significant signals.
Therefore, we performed a source-by-source check and removed signals corresponding to 
activity in known $\gamma$-ray blazars.

\section{Search for Galactic transients}

In this section, we present a search for both non-repeating and repeating transient \gray{} sources in the Galactic plane using the VSSTW method.

\subsection{Transient \gray{} sources in the Galactic plane}

\begin{table*}
\centering \caption{The list of transient \gray{} signals obtained
from the performed VSSTW analysis. The second and third
columns show the Right Ascension and the Declination of a transient \gray{} source. 
The fourth and fifth columns show the start date and the length
of a high-\gray{}-flux state. The six and seventh columns show the local and
global significances at which the high flux state is detected.
The eighth and ninth
columns show the name and class of a \gray{} source associated with a transient signal. 
The tenth column shows the distance between the positions of transient and associated sources.}
\begin{tabular}{ | c | c | c | c | c | c | c | c | c | c|}
\hline
Identifier &  R.A. & Dec.  &  \gray{} bright  & Length  & Local. &  Global & Assoc. source & Class & Distance \\
 & (deg)       &  (deg)         &  from yr/m/d          & (week)  & signif. ($\sigma$)  & signif. ($\sigma$)   & &  & (deg)        \\
& &  &  &   &  &  & & & \\
\hline
G01 & 159.457 & -59.489 & 2018/03/28 & 4 & 31.8 & 31.4 & V906 Car & nova & 0.23\\
G02 & 315.371 & 45.919 &  2010/03/03 & 2 & 17.3 & 16.5 & V407 Cyg & nova & 0.19 \\
G03 & 305.080 & 40.384 & 2009/07/15 & 118 & 13.6 & 12.4 & PSR J2021+4026 & plsr & 0.24\\
\textbf{G04} & 308.026 & 40.952 & 2020/04/15 & 26 & 13.5 & 12.4 & Cygnus X-3 & $\mu$qsr & 0.06\\
G05 & 32.939 & 64.366 & 2017/03/08 & 5 & 13.2 & 12.1 & N08 & bcu? & 0.80\\
G06 & 40.149 & 61.180 & 2014/02/12 & 134 & 12.1 & 9.8 & LS I +61${^{\circ}}$303 & $\gamma$bin & 0.05\\
\textbf{G07} & 267.803 & -32.734 & 2012/06/06 & 2 & 10.5 & 9.1 & V1324 Sco & nova & 0.13\\
\textbf{G08} & 195.518 & -63.892 & 2017/10/25 & 5 & 10.3 & 8.9 & PSR B1259-63 & $\gamma$bin & 0.10\\
G09 & 99.722 & 5.771 & 2012/06/13 & 2 & 8.8 & 7.1 & V959 Mon & nova & 0.23\\
G10 & 70.866 & 47.326 & 2018/04/18 & 2 & 8.6 & 6.9 & V392 Per & nova & 0.04\\
G11 & 208.521 & -58.907 & 2013/11/27 & 6 & 7.6  & 5.6 & V1369 Cen & nova & 0.26\\
\textbf{G12} & 272.510 & -27.378 & 2016/10/19 & 3 & 7.0 & 4.8 & V5855 Sgr & nova & 0.15\\
\hline
G13 & 84.822 & 21.865 & 2011/04/06 & 1 & 6.9 & 4.6 & Crab & plsr & 1.11\\
\textbf{G14} & 171.385 & -65.746 & 2017/12/27 & 6 & 6.5 & 4.0 & V357 Mus & nova & 0.24\\
\hline
\end{tabular}
\label{Tab}
\end{table*}

We present the results in Table \ref{Tab} which contains the list of transient 
\gray{} sources whose high flux states are revealed by the performed VSSTW analysis at local and global significance levels above 7$\sigma$ 
and $\approx5\sigma$, respectively. Twelve transient \gray{} signals shown above the horizontal line in Table \ref{Tab} 
are at local significance levels higher than 7$\sigma$. To these signals, we added two signals at lower local significances of
$6.9\sigma$ and $6.5\sigma$, given their robust identification with V357 Mus and the Crab pulsar nebula, respectively. 
The identifiers of \gray{} signals that are not present in the paper by \citet[][]{PMV2021} are shown
in bold. Among the $\gamma$-ray signals that occurred during coincident time intervals
and at neighbouring positions in the sky, we listed the one with the highest significance in Table \ref{Tab}.

\begin{figure}
\centering
\includegraphics[angle=0, width=.5\textwidth]{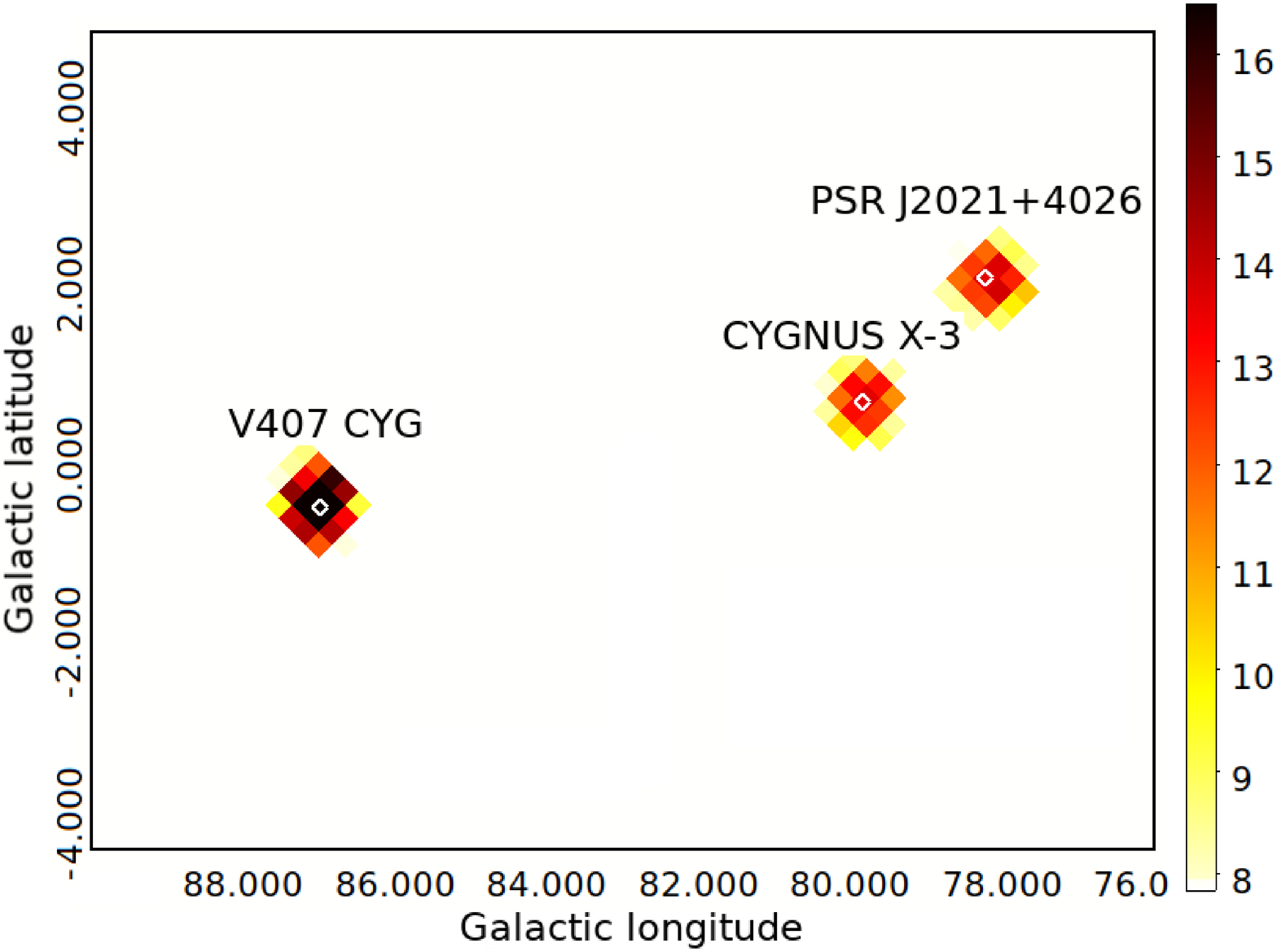}
\caption{The significance map of $\gamma$-ray transient emission in $\sigma$ showing the microquasar Cygnus X-3, the nova V407 Cyg, and the pulsar PSR J2021+4026.}
\label{Fig1}
\end{figure}

\textbf{Gamma-ray binaries.}
Table \ref{Tab} contains two \gray{} binaries, LS I +61$^{\circ}$303 and PSR B1259-63. The former binary with an orbital period of 26.5 days is known for its long-term $\gamma$-ray variability associated with a superorbital period of 1667 days \citep[][]{Gregory2002}. 
The start of a high-\gray{}-flux state on the date indicated in this Table is in agreement with that reported by \citet[][]{PMV2021} and corresponds to the maximum after 1667 days since the previous maximum reported by
\citet[][]{superorb}. The other binary reported in Table \ref{Tab}, PSR B1259-63, has an orbital period of 1237 days.   
Its high-\gray{}-flux state started in October 2017 following the periastron passage on 22 September
2017. This result confirms that the GeV flaring state only started 40 days after the 2017 periastron and lasted approximately 50 days \citep[][]{Johnson}. The GeV \gray{} emission associated with the 2017 periastron passage of PSR B1259-63 showed a very different behaviour than those in the 2010 and 2014 flaring events \citep[e.g.,][]{Johnson}.
The spectrum reported by \citet[][]{Johnson} exhibits a cut-off at $\simeq$800 MeV during the flare 
in 2017. The inclusion of \textit{Fermi}-LAT data between energies from 500 MeV to 800 MeV in the VSSTW search increases 
the local significance of this post-periastron flare from 6.2$\sigma$ to 10.3$\sigma$ explaining why
this high flux state was not reported by \citet[][]{PMV2021}.

\begin{figure}
\centering
\includegraphics[angle=0, width=.5\textwidth]{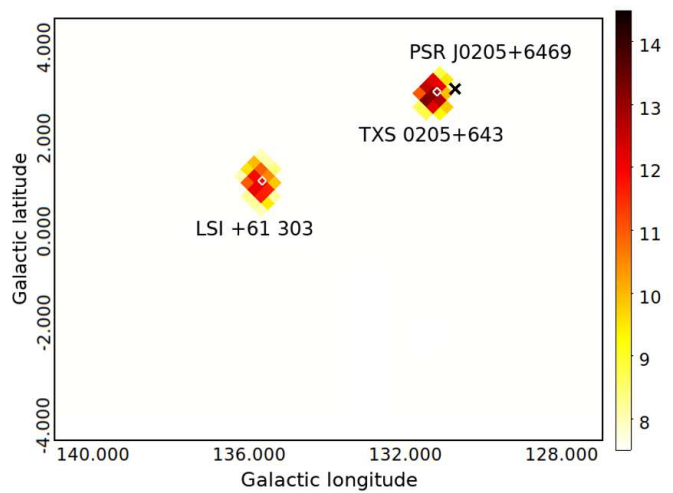}
\caption{The significance map of $\gamma$-ray transient emission in $\sigma$ showing the binary LS I +61$^{\circ}$303, the $\gamma$-ray source N08/G05, and the source, TXS 0205+643.}
\label{Fig2}
\end{figure}

\textbf{Novae.} Our search resulted in the detection of 8 novae, V906 Car, V407 Cyg, V1324 Sco,
V959 Mon, V392 Per, V1369 Cen, V5855 Sgr, and V357 Mus. Four of these 8 novae, namely V407 Cyg, V1324 Sco,
V959 Mon, and V1369 Cen, were included in the second \textit{Fermi} all-sky variability analysis catalogue \citep[2FAV;][]{Fermi17}. Three novae, V906 Car, V5855 Sgr and V357 Mus, 
happened in March 2018, October 2016, and December 2018, respectively, are not present in the 2FAV catalogue, which covers \textit{Fermi}-LAT observations until January 2016. 
Five of these 8 novae, namely V906 Car, V407 Cyg, V959 Mon, V392 Per, and V1369 Cen, 
were confirmed using the VSSTW method by \citet[][]{PMV2021}.
In that paper, the other two confirmed novae, V339 Del and V5856 Sgr, are at higher Galactic latitudes than those covered by the current search. In addition to the novae reported by
\citet[][]{PMV2021}, we confirmed V1324 Sco, V5855 Sgr, and V357 Mus and increased the sample of novae detected with this
analysis technique by 60\%. 
The novae, V5855 Sgr and V357 Mus, have fainter $\gamma$-ray fluxes
\citep[][]{Gordon2021} than the other novae in Table \ref{Tab}. Although their global significance levels are
below 5$\sigma$, these levels, $4.8\sigma$ and $4.0\sigma$, are sufficiently high to detect these signals and for their identification with these novae owing to the localization in both time and the sky. 

Two novae at low Galactic latitudes detected in \textit{Fermi}-LAT data, but not present in Table \ref{Tab}
are V549 Vel \citep[][]{LiKwanLok2020} and V1707 Sco \citep[][]{Sco2019}, the faintest novae in $\gamma$ rays.

\textbf{Microquasars.} We detected a microquasar, Cygnus X-3, using the VSSTW method.
Given that the detected high flux state is in 2020 and beyond the time interval covered by 
\citet[][]{PMV2021}, we present a detailed analysis of this signal in Section 4.
Figure \ref{Fig1} shows the significance map including transient signals from three $\gamma$-ray sources of different classes,
namely the microquasar Cygnus X-3, the nova V407 Cyg, and the pulsar PSR J2021+4026.

\textbf{Pulsars, pulsar nebula, and the source G05.}
Four \gray{} signals in Table \ref{Tab} are localized in the directions of pulsars, 
including PSR J2021+4026 (G03), PSR J0205+6449 (G05), and the Crab pulsar (G13). 
The signal, G03, is at a distance from the near pulsar of $0\fdg24$ that corresponds to the mean spacing in the HEALPix grid of resolution $N_\mathrm{side}$ = 256 and its association is robust.
The pulsar, PSR J2021+4026, is indeed a variable \gray{} pulsar whose flux decreased on 2011 October 16, see 
\citet[][]{Allafort2013, PMV2021}. The start date and the length of a high-\gray{}-flux state of PSR J2021+4026 
reported in Table \ref{Tab} are compatible with those reported by \citet[][]{PMV2021}. 
The Crab pulsar nebula $\gamma$-ray `superflare' that occurred on 2011 April 12 \citep[][]{Buehler} was confirmed
as a transient \gray{} signal at a local significance above 8$\sigma$ by means of the VSSTW method by \citet[][]{PMV2021}. 
The Crab pulsar, located at a Galactic latitude of $-5\fdg8$, is at a large distance ($\simeq1\fdg1$) from 
the associated pixel (in Table \ref{Tab}) because of the separation of Crab from the region used in this search, see also the
case of PKS 1830-21 described above. This separation also explains a lower significance of the Crab nebula in 
Table \ref{Tab} compared to that previously reported.
The other transient \gray{} source at a significant distance from the associated pixel, N08, is near PSR J0205+6449 that is a 65-millisecond young rotation-powered pulsar. At the position of this pulsar
the local significance is $7.8\sigma$, while at the associated pixel the 
local significance is $13.2\sigma$. 
Figure \ref{Fig2} shows the significance map obtained from the VSSTW
analysis with the positions of PSR J0205+6449 and a radio source, TXS 0205+643, and illustrates this fact. The position of PSR J0205+6449 is shown by a cross.
This difference in significances is suggestive that another $\gamma$-ray source can 
be responsible for this transient \gray{} signal. The \textit{Fermi}-LAT count map shown in Figure 4 of \citet[][]{PMV2021}
also indicates a similar offset from the position of PSR J0205+6449 to the east. The flare of the source N08 was
in 2017 and the \gray{} flux was 3.7 times higher \citep[][]{PMV2021} than that of the nearest source, PSR J0205+6449, in the \textit{Fermi}-LAT 10-year source catalogue (4FGL-DR2). When our manuscript was in preparation, the 
\textit{Fermi}-LAT collaboration released a new catalogue (4FGL-DR3) that is based on 12 years of data
\citep[][]{cat12}. In the 4FGL-DR3 catalogue, a new source, namely 4FGL J0209.7+6437, to the east 
of PSR J0205+6449 is added and identified with a blazar of uncertain type, TXS 0205+643 or NVSS J020935+643725. 

Figure \ref{Fig2} also shows the \gray{} binary LS I +61$^{\circ}$303, which is one of the $\gamma$-ray sources expected to exhibit repeating signals on the time scale of years.

\subsection{Repeating high-\gray{}-flux transient signals}

We presented above the results obtained from the VSSTW analysis applied to 13 years of \textit{Fermi}-LAT data with the purpose of identifying the strongest $\gamma$-ray flare for each source. 
There exist, however, $\gamma$-ray sources showing modulated emission with a period longer than 3 years, such as PSR B1259-63 and $\eta$ Carinae \citep[e.g.,][]{Johnson, MartiDevesa2021}. To search for repeating flares from such sources during shorter time intervals than 13 years, we divided the entire data into four equal time intervals of 170 weeks each (that is about three and a quarter years) and performed a search for $\gamma$-ray transient signals in each of these time intervals by means of the VSSTW method. 
We repeated the analysis of data divided into 170-week intervals selecting data with a 85-week shift for improving the sensitivity for sources whose flaring activity is at the edges of the time intervals.
68476 positions in the Galactic plane used in this search are the same as those used in the previous section.

The performed search resulted in 8 additional flaring events, two of them are during the time interval that is from August 2008 to November 2011, another one is during the time interval that is from June 2013 to September 2016, the next two are during the time interval that is from September 2016 to January 2020, and
the remaining three are in the time interval that is from May 2018 to July 2021. These 8 events are from four $\gamma$-ray sources, namely PSR B1259-63 (two events), LS I +61$^{\circ}$303 (two events), PSR J2021+4026 (three events), and Cygnus X-3.

\begin{figure}
\centering
\includegraphics[angle=0, width=.5\textwidth]{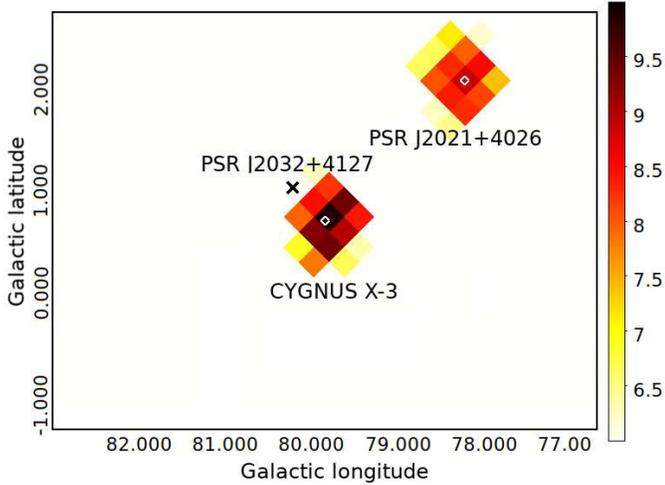}
\caption{The significance map of $\gamma$-ray transient emission in $\sigma$ showing the Cygnus region and corresponding to the 170-week data subset from May 2018 to July 2021.}
\label{Fig3}
\end{figure}

\begin{figure*}
\centering
\includegraphics[angle=0, width=.95\textwidth]{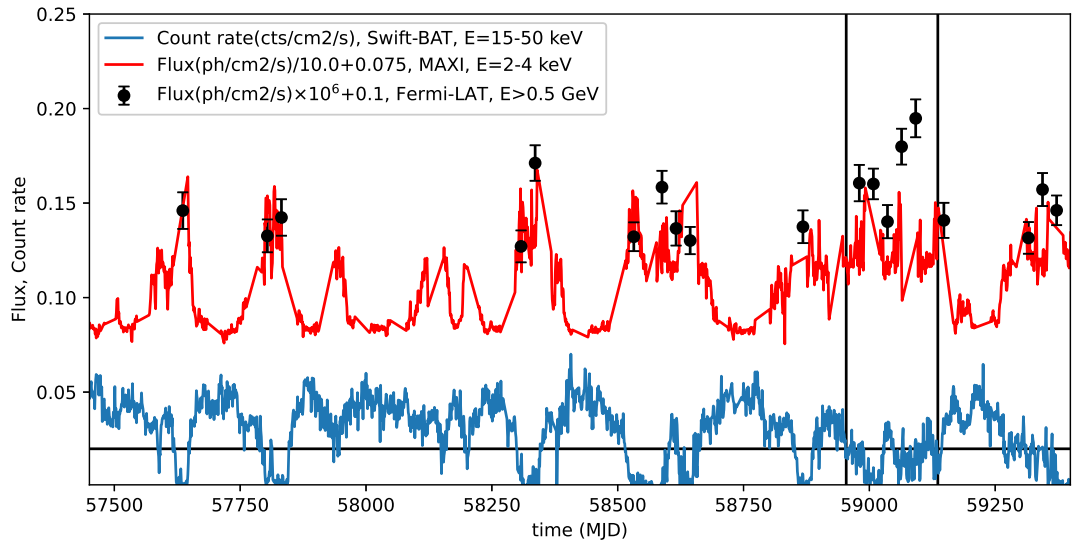}
\caption{\textit{Fermi}-LAT, MAXI, and \textit{Swift}-BAT light curves of Cygnus X-3 showing data points
from April 2016 to July 2021.}
\label{Fig4}
\end{figure*}

\textbf{PSR B1259-63.} Two high-\gray{}-flux events of PSR B1259-63 started in January 2011 and March 2021 and lasted for 3 week and 8 weeks, respectively. The local significances of these
transient $\gamma$-ray signals are 7.1$\sigma$ and 6.4$\sigma$. 
These high-\gray{}-flux events are associated with the periastron passages of PSR B1259-63 in December 2010 \citep[][]{psr1259per2010} and February 2021 \citep[][]{psr1259per2021_1, psr1259per2021_2}, respectively. 

\textbf{LS I +61$^{\circ}$303.} Two high-\gray{}-flux events of LS I +61$^{\circ}$303 started in March 2009 and April 2019 and lasted for a few years (about 136 and 95 weeks, respectively). These events are at local significances of 10.2$\sigma$ and 8.5$\sigma$ and
most likely associated with superorbital modulation and correspond to
the preceding \citep[][]{superorb} and succeeding high-$\gamma$-ray-flux states with respect to those reported in Table \ref{Tab}.

The selected 170-week intervals partially or significantly overlap with one of these three high-flux intervals and the VSSTW method applied to these intervals provides evidence for superorbital modulation. Meanwhile, the VSSTW method does not show a high-flux state lasting 4 weeks or less in duration corresponding to the orbital modulation with the period of 26.5 days. This is most likely due to the fact of each 170-week interval comprises many (45) lengths equal to the period. The orbital period modulation of LS I +61$^{\circ}$303 in $\gamma$ rays was established by other methods \cite[see, e.g.,][]{PM2017}.

\textbf{PSR J2021+4026.} This pulsar experienced a decrease in flux near 2011 October 16 \citep[][]{Allafort2013} 
that corresponds to the end of the high-\gray{}-flux state reported in Table \ref{Tab}. In addition to that dimming event in
2011, further $\gamma$-ray state changes of PSR J2021+4026 in 2014 and 2018 were reported by \citet[][]{Fiori}.
We found the flux changes in December 2014 and February 2018 with
significances of 10.3 $\sigma$ and 10.4 $\sigma$ each confirming the result reported by \citet[][]{Fiori}.
They also reported that PSR J2021+4026 was in its low flux state from 2018 February 2 to 2020 May 26. 
The performed VSSTW analysis allowed us to reveal a new high-\gray{}-flux event at a significance level of 8.5$\sigma$ started in June 2020. 

Figure \ref{Fig3} shows the significance map centered on the position of Cygnus X-3, obtained from the VSSTW analysis, and based on the data
accumulated during the most recent time interval. Two transient signals from Cygnus X-3 and PSR J2021+4026 are significantly detected, while no transient signal is detected from the pulsar PSR J2032+4127 which is located at a distance of $0\fdg50$ from Cygnus X-3. PSR J2032+4127 is a binary with a period of 45-50 years \citep[e.g.,][]{psrj2032_1}. The non-detection of a transient GeV signal from PSR J2032+4127 is
consistent with no change in flux at GeV energies during the 2017 periastron passage of PSR J2032+4127 and also between August 2008 and December 2019 reported by \citet[][]{psrj2032_2}. We performed a sanity
check by comparing the fluxes from a test position located at the same distance from Cygnus X-3 as PSR J2021+4026 (and at a distance of 3\fdg1 from PSR J2021+4026)
before and during the \gray{} bright state of Cygnus X-3 in 2020  and found that the variation in \gray{} flux at that test position is significantly smaller than the change in flux of PSR J2021+4026 in June 2020, ensuring that the \gray{} flux changes of Cygnus X-3 and PSR J2021+4026 in 2020 are different events.

\textbf{Cygnus X-3.} In addition to the $\gamma$-ray flaring event in mid-2020 and lasting for about six months as reported in the next section, we found another flaring event from Cygnus X-3 at a statistical significance of 7.0$\sigma$ when we applied the analysis to the time interval that is from September 2016 to January 2020. This additional flaring event started in the beginning of August 2018 and lasted for two weeks.

\section{Flare of Cygnus X-3 in mid-2020}

In this section, we present the results of \textit{Fermi}-LAT observations of Cygnus X-3 
during the first 13 years of the \textit{Fermi} mission and a search for modulated \gray{} emission
from Cygnus X-3 during the flare in 2020.

\subsection{Likelihood analysis with \texttt{FERMITOOLS}}

We binned the data into 181 28-day intervals in order to produce a light curve and
analysed \textit{Fermi}-LAT data collected during these time intervals using the binned 
maximum likelihood mode of \texttt{gtlike}, which is part of \texttt{FERMITOOLS}. 
We employed the TS \citep[][]{Mattox} to evaluate the significance of the \gray{} emission
coming from the source located at the position of Cygnus X-3 during each of these 28-day intervals.
Along with Cygnus X-3, the model includes 4FGL sources within a region of $20^{\circ}$ radius around Cygnus X-3. 
We took their spectral shapes from the 4FGL catalogue \citep[][]{Fermi20}. 
We selected the energy range for this analysis from 0.5 GeV to 300 GeV with 25 logarithmic energy
bins and used spatial binning with a pixel size of 0\fdg05. We allowed the power-law normalisation and 
photon index of Cygnus X-3 
and the normalisations of bright 4FGL \gray{} sources, in particular PSR J2021+4026, PSR J2032+4127, and 
the Cygnus cocoon, to vary, while keeping the normalisations of other 4FGL sources fixed.
We adopted a background model that includes components describing the diffuse Galactic and isotropic
\gray{} emission. We used the standard templates \texttt{gll\_iem\_v07.fits} and
\texttt{iso\_P8R3\_SOURCE\_V2\_v1.txt} for the Galactic diffuse and isotropic components,
respectively, and allowed their normalisations to vary as well. 

We computed the light curve and found that 23 of the 181 data points are with TS values greater than 16, 
each corresponding to a detection of the source at a significance level greater than $4\sigma$. The highest, second-, fourth-, and seventh-highest TS values are 168, 115, 95, and 67 (that is $13.0\sigma$, $10.7\sigma$, $9.7\sigma$, and $8.2\sigma$), respectively, and correspond to four 28-day intervals in the course of the flaring event detected by means of the VSSTW search in Section 2.3.
Figure \ref{Fig4} shows the obtained light curve including only the \gray{} data points with $TS>16$.  The time interval of the flaring event in 2020 lies between two vertical lines in this figure.  
Figure \ref{Fig4} also shows the soft (2-4 keV) and hard (15-50 keV) X-ray light curves based on
data points from MAXI \citep[][]{maxi} and Swift-BAT \citep[][]{bat} and taken from the webpages, \burl{http://maxi.riken.jp/star_data/J2032+409/J2032+409.html} and \burl{https://swift.gsfc.nasa.gov/results/transients/CygX-3/}, respectively.
It is acknowledged that Cygnus X-3 exhibits an overall anticorrelation between soft and hard X-ray fluxes \citep[e.g.,][]{Szostek} and also an overall anticorrelation between hard X-ray and \gray{} emission \citep[][]{CygX3Sci, CygX3Nat}. The light curves in Figure \ref{Fig4} shows that \gray{} flaring events of Cygnus X-3 correspond to deep local minimums of its hard X-ray emission \citep[with a \textit{Swift}-BAT count rate of $\leq$0.02 cts cm$^{-2}$ s$^{-1}$ as suggested by][]{Piano2012}. The horizontal line shown in Figure \ref{Fig4} corresponds to this threshold 
\textit{Swift}-BAT count rate and indicates that Cygnus X-3 was in a soft X-ray state during the \gray{} flaring event in 2020. For the sake of illustration, we selected the start date of 2016-04-22 (MJD 57500) while showing these light curves. 

We also performed a likelihood analysis of \textit{Fermi}-LAT data accumulated during 5 consecutive 28-day intervals during the flaring event in 2020. The corresponding data points are shown in Figure \ref{Fig4}
between two vertical lines. Each of these intervals allows a significant detection of \gray{} emission from Cygnus X-3. The joint analysis of these 5 data sets resulted in the TS value for Cygnus X-3 of 448, which corresponds to $21\sigma$ significance. The likelihood maximisation yields a photon index of $\Gamma=3.12\pm0.10$ and an integral flux of $(6.74\pm0.41)\times10^{-8}$ ph cm$^{-2}$ s$^{-1}$ above 0.5 GeV. 

In addition to the strongest \gray{} flaring event,
Figure \ref{Fig4} shows that there are other significantly detected \gray{} flaring events observed 
during the intervals of faint hard and powerful soft X-ray emission. The performed analysis resulted in 18 additional detections in 28-day intervals. 
Five of these 18 \gray{} detections occurred before 2016-04-22. 
The results of \textit{Fermi}-LAT 
observations corresponding to the first four of them and the fifth one were reported by \citet[][]{CygX3Sci} and \citet[][]{Corbel2012}, respectively.  
The data points provided by 13 detections in 28-day intervals occurring after 2016-04-22 are shown in Figure \ref{Fig4}.  
These 13 data points correspond to 6 soft X-ray states of Cygnus X-3.
The \gray{} flaring events corresponding to these soft X-ray states were in August - September 2016, February - April 2017, June - August 2018, February  - June 2019, January - February 2020, and March - May 2021. 
All these 6 \gray{} flaring events of Cygnus X-3 have been reported to be present in AGILE data and
the first two of them in \textit{Fermi}-LAT data, see Astronomer’s Telegrams\footnote{\burl{https://www.astronomerstelegram.org/}; These Astronomer's telegrams include 
ATel \#9429 (AGILE) and ATel \#9502 (\textit{Fermi}-LAT) for the first of these 6 intervals, ATels \#10109 and \#10243 (both \textit{Fermi}-LAT) and ATels \#10138 and \#10179 (both AGILE)
for the second interval, ATels \#11804 and \#11814 (both AGILE) for the third interval, Atels \#12677, \#12678, and 
\#12894 (all AGILE) for the fourth interval, ATels \#13423 and \#13458 (both AGILE) for the fifth interval, and in ATels \#14662 and \#14780 (both AGILE) for the sixth interval. See, Atel \#15009 for $\gamma$-ray activity detected by AGILE in October 2021 (i.e., started after the time interval included in this paper).}. 
In this Section, the last four of these \gray{} flaring events 
are for the first time reported to be present in \textit{Fermi}-LAT data. 
Note that the \gray{} flare revealed by means of the
VSSTW analysis, that is G04 from Table \ref{Tab}, has not been reported as Astronomer's Telegrams by the \textit{Fermi}-LAT or AGILE collaborations.

\subsection{4.8-hour gamma-ray pulsations of Cygnus X-3}

To assess the significance of \gray{} modulation during the flaring event in 2020,
we binned \grays{} into 1000-second time bins and performed a Poisson maximum likelihood analysis of data extracted from the $0\fdg35$ radius aperture around Cygnus X-3. We chose the aperture size smaller than the distance from Cygnus X-3 to PSR J2032+4127. 
We modelled the expected number of counts in a bin centered on time, $t_\mathrm{i}$, as
$N(t_\mathrm{i})=\epsilon_{\mathrm{i}}\times\left(F_{0}+F_{1}\times\left(1+\cos\left(2\pi\left(t_{\mathrm{i}}-T_{0}\right)/P+\delta\right)\right)\right)$, where $\epsilon_\mathrm{i}$, $T_0$, $P$, and $\delta$ are the exposure, start time, period, and phase. Compared to the equation in the Supporting Online Material for \citet[][]{CygX3Sci}, we included $2\pi$ (its absence there appears to be a misprint) and renormalised $F_{1}$ by adding 1 to the cosine function making the modelled flux always non-negative.\\

\begin{figure}
\centering
\includegraphics[angle=0, width=.5\textwidth]{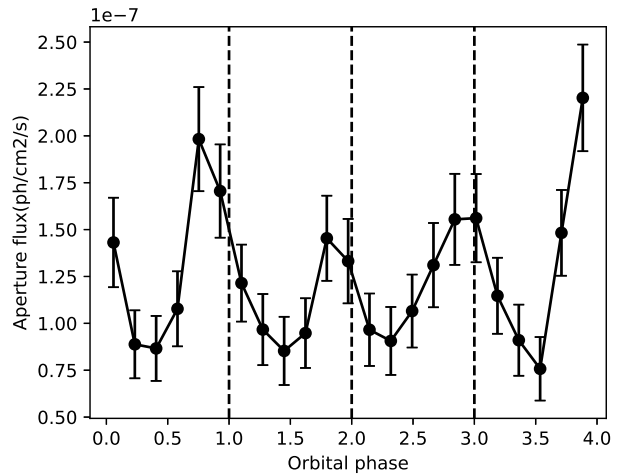}
\caption{The light curve folded on the time interval corresponding to the four orbital periods for the \textit{Fermi}-LAT data in the time range JD 2458967.2871--2459107.1494. The length of 1 unit of phase 
is equal to the orbital period and the compact object is behind the Wolf-Rayet star at
phase=0 (superior conjuction).}
\label{Fig5}
\end{figure}

When considering emission from the entire flaring event consisting of 5 28-day intervals, we found that a modulated flux model improves the log likelihood over a constant flux model by $\Delta\ln{L}=31.0$. Twice the difference in log likelihood follows a $\chi^2$ distribution with 3 degrees of freedom
\citep[additional degrees of freedom are $F_{1}$, $P$, and $\delta$, see][]{CygX3Sci}. 
This change in the log likelihood corresponds to a significance level of $7.3\sigma$. 
We derived the period using a profile likelihood method. The best-fit period is $P=4.79298\pm0.00045$ hours
and in reasonable agreement with the period, $P=4.7917\pm0.0011$ hours, obtained by \citet[][]{CygX3Sci} from the active states of Cygnus X-3 in 2008-2009 and with the orbital period of Cygnus X-3. We computed the statistical uncertainties using the likelihood profile for which we changed the period, $P$, until the maximum likelihood decreases by 0.5 in logarithm. We report the statistical uncertainties, however they should be treated with caution, since systematic uncertainties are also present due to the assumed cosine-shape of the modulated signal and the assumption of a constant flux amplitude during the modulated signal. The uncertainty due to the latter assumption can be reduced by dividing the entire interval into similar flux intervals. We show the folded $\gamma$-ray light curve obtained from the \textit{Fermi}-LAT data accumulated during the active state of Cygnus X-3 in mid-2020 in Figure \ref{Fig5}. 
The maximum $\gamma$-ray flux in mid-2020 occurs just before superior conjunction. Applying a sinusoidal model to the phase-folded data, we estimated that the phase of maximum of $\gamma$-ray emission is at $\phi=0.84$.
\citet[][]{CygX3Sci} reported that the maximum $\gamma$-ray flux during the active state in 
2008 and 2009 also occurs just before superior conjunction at $\phi=0.88$ and $\phi=0.76$, respectively. Therefore, $\gamma$-ray emission is produced when energetic particles propagating from the compact object, as seen behind the Wolf-Rayet star.
The presence of modulated \gray{} emission during the flaring event in 2020 strengthens the previous
conclusion about modulation drawn from the first two years of \textit{Fermi}-LAT observations of Cygnus X-3. 

\begin{table}
\centering \caption{The start and end times of the five time intervals during the \gray{} flare
of Cygnus X-3 in 2020 shown in \textit{Fermi} mission elapsed time (MET; in seconds).}
\begin{tabular}{ | c | c | c | }
\hline
Interval \# &  Start & End \\
\hline
1 & 609695017 & 612114217 \\
2 & 612114217 & 614533417 \\
3 & 614533417 & 616952617 \\
4 & 616952617 & 619371817 \\
5 & 619371817 & 621791017 \\
\hline
\end{tabular}
\label{Tab2}
\end{table}

We also searched for the presence of periodic emission from the source for each of the 5 28-day intervals of the flaring event. The start and end times of these intervals are listed in Table \ref{Tab2}.
We found that the source emits periodic \gray{} emission during the fifth 28-day interval, when the source is the brightest. There is some weak evidence for periodic emission during three of the other four 28-day intervals. 
We found that a modulated flux model improves the log likelihood over a constant model by $5.1$, $5.9$, $2.8$, $4.3$, and $15.9$ for the first to the fifth 28-day intervals. The corresponding significances for the presence of periodic $\gamma$-ray emission are $2.4\sigma$, $2.6\sigma$, $1.5\sigma$, $2.1\sigma$, and $5.0\sigma$.
The best-fit period corresponding to the \gray{} brightest 28-day interval is $P=4.7972\pm0.0033$ hours.
The period reported by \citet[][]{CygX3Sci} is well within the 90\% confidence interval and the orbital period, $P=4.7926$ hours, extrapolated from \citet[][]{Singh} is in better agreement. 
We performed a combined analysis of the first four 28-day intervals. The combined analysis yields $\Delta\ln{L}=19.1$ corresponding to a $5.5\sigma$ significance level. The period obtained from the combined analysis is $P=4.79388\pm0.00083$ hours and in agreement with the period derived from the entire flaring event and the orbital period. We also compared the flux amplitudes of modulated emission, $F_{1}$,
corresponding to the combined four and fifth 28-day intervals and found that the amplitude, $F_{1}$, is
by a factor of about 2 higher for the fifth interval. 
This fact supports that the flux increase during the fifth interval was due to the modulated emission.
Both the consistency of the derived period with the orbital period and the fact that each of the studied time intervals spans many cycles of the period \citep[see,][for the false positive rate of few-cycle periodicities]{Vaughan2016} strongly support the reliability of the obtained results.

\subsection{Outlook}

The detection of modulated $\gamma$-ray emission during the flare from Cygnus X-3 in 2020 
opens new opportunities to study the conditions at which $\gamma$-ray activity in this microquasar 
is produced. Searches for modulated emission during the other 6 flaring events shown
in Figure \ref{Fig4} is required. 
To increase the statistic one can also perform a stacking analysis
\citep[see, e.g.,][]{occ} of the \textit{Fermi}-LAT observations during these flaring events.
Another way to increase the statistic is to use a larger aperture
and to subtract the contribution of PSR J2032+4127 to the $\gamma$-ray signal using the pulsar gating technique
\citep[see][]{CygX3Sci}. The pulsar gating technique can be useful to further 
boost the significance of the detected modulation during the 2020 flaring event allowing a
more detailed study.

Multi-wavelength studies of Cygnus X-3 from the lowest to the highest frequency during 
different X-ray states and during $\gamma$-ray activity in mid-2020 can provide further information about this binary system. Cygnus X-3 exhibits occasional giant radio flares as intense as a few thousand times the quiescent emission level in the radio band, first seen by \citet[][]{Gregory1972}. Giant radio flares in 2011, 2016, and 2017 marked the transitions between different X-ray states, accompanied by rising non-thermal hard X-ray emission \citep[][]{Corbel2012, Trushkin2017}. Most recently, a giant radio flare was observed in February 2020 \citep[][]{Green2020}. Examples
of major and minor flares and comparison of their physical parameters are given by \citet[][]{Spencer2022}. Meanwhile, comparison of the results obtained with TeV $\gamma$-ray observations between 2006 and 2011 by the MAGIC and VERITAS telescopes \citep[][]{MAGIC2010, VERITAS2013} with those by the SHALON telescope \citep[][]{Sinitsyna2018} can provide insights if TeV $\gamma$-ray emission is created under specific conditions.

\section{Summary}

We performed a VSSTW analysis with the purpose of searching for transient \gray{} signals in the Galactic plane using 13 years of the \fermilat{} data. Compared to the previous search by means of this technique by \citet[][]{PMV2021}, besides using more years of data we restored full coverage, used finer pixelization of the Galactic plane, and also broadened the energy range down to 0.5 GeV. The detected sources are listed in Table \ref{Tab}.
This refined search increased the number of transient \gray{} sources in the Galactic plane by more than 50\% from the number reported in the previously published search.
Among these sources are a \gray{} binary (PSR B1259-63), three more novae (V1324 Sco, V5855 Sgr, and V357 Mus), and, in particular, a microquasar, Cygnus X-3. The \gray{} binary PSR B1259-63 is a soft \gray{} source and the inclusion of lower energy \grays{} was crucial for its detection. The nova, V1324 Sco, is a strong transient \gray{} source that had avoided detection in the previous VSSTW search due to sparser coverage of the Galactic plane. The novae, V5855 Sgr and V357 Mus, are faint in \grays{} and the increase of aperture area by 2 times (and therefore statistics) is a relevant factor for their detection. The refined search also allowed us to study in more detail the transient signal, N08 (or G05 in Table \ref{Tab}), that is near PSR J0205+6449. 
The significance map shown in Figure \ref{Fig2} illustrates that the position of the highest significance is at an offset from PSR J0205+6449 to the east. This fact is suggestive that another \gray{} source can be responsible for this transient signal.

We also performed a VSSTW analysis of data subsets consisting of 170 weeks of data each and searched for repeating high-\gray{}-flux transient signals. This search reveals repeating signals from PSR B1259-63, LS I +61$^{\circ}$303, PSR J2021+4026, and Cygnus X-3 additional to those in Table \ref{Tab}. The repeating signals from PSR B1259-63 are identified with periastron passages of the binary system and with a period of about 1237 days. The repeating signals from LS I +61$^{\circ}$303 are most likely associated with 1667 day superorbital modulation 
of the binary system. The latest of these three high states of LS I +61$^{\circ}$303 started in 2019 and its detection had not yet been reported at GeV energies. The search showed that in addition to the dimming event in 2011 listed in Table \ref{Tab}, the pulsar, PSR J2021+4026, experienced two flaring events, one was between 2014 and 2018 and the other started in 2020. The former had recently been reported by \citet[][]{Fiori} and the latter had not been reported. The second flaring event in Cygnus X-3 lasted for two weeks corresponds to a soft X-ray state in August 2018.

The VSSTW analysis revealed a high-\gray{}-flux state of the microquasar, Cygnus X-3. This flaring event happened in 2020. By comparing the light curves in the \gray{}, soft X-ray, and hard X-ray bands, we found that Cygnus X-3 was in a soft X-ray state confirming the previously noted trend \cite[][]{CygX3Sci, CygX3Nat}. We performed a binned likelihood analysis and found that the \gray{} spectrum of Cygnus X-3 corresponding to this high-\gray{}-flux state is with a relatively soft photon index of $3.1$ and that the integral \gray{} flux above 0.5 GeV is as high as that reported by \citet[][]{CygX3Sci} for the flaring events in 2008 and 2009. Given the high \gray{} flux and the long duration of the \gray{} flaring event in 2020, this event provided us with an opportunity to search for orbital modulation of the \gray{} emission. We found that the modulation during this flaring event has high significance and the best-fit period of 4.793 hours. The obtained period is in agreement with that derived from \textit{Fermi}-LAT observations of the previous flaring events from Cygnus X-3 \citep[][]{CygX3Sci} and the orbital period of this binary system. We also found that the phase of maximum $\gamma$-ray emission in mid-2020 is around superior conjunction. This conclusion suggests that $\gamma$-ray emission seen in 2008-2009
and in mid-2020 are likely to be produced by the same mechanism.

\

\section{Acknowledgements}

We are grateful to Jacco Vink for valuable suggestions and discussions
and thank the referee for the constructive comments that helped us to improve the manuscript.

Computations were performed on the computational facilities belonging to the ALMA Regional Center Taiwan, 
Academia Sinica, Taiwan. D.P. is supported by funding from the European
Union's Horizon 2020 research and innovation program
under grant agreement No. 101004131 (SHARP).

\section{Data availability}

\textit{Fermi}-LAT data analysed in this paper are publicly
distributed by the LAT team and can be downloaded from the LAT Data
Server. The python code used to produce the results of the paper is publicly available.
The significance maps obtained from the VSSTW analysis applied to 
the entire data set and each of the analysed subsets of data are publicly
available in the fits format at \burl{https://zenodo.org/record/7348674}.

\section{A note in proof}

After acceptance of our manuscript for publication, we became aware 
of the paper by \citet{Zdziarski2018}, who provided an extensive
study of Cygnus X-3 using \textit{Fermi}-LAT data collected between 2008
August and 2017 August. Note that the two flaring events of Cygnus X-3 detected 
with the VSSTW method are after that time range and that other newly reported flaring events 
of Cygnus X-3 in Section 4.1 are indeed reported for the first time.

\end{document}